# EXOPLANET TRANSITS REGISTERED AT THE UNIVERSIDAD DE MONTERREY OBSERVATORY. PART I: HAT-P-12b, HAT-P-13b, HAT-P-16b, HAT-P-23b AND WASP-10b

BY

PEDRO V. SADA [1]   &   FELIPE G. RAMÓN-FOX [2]

[1] Universidad de Monterrey, Departamento de Física y Matemáticas, Av. I. Morones Prieto 4500 Pte., San Pedro Garza García, Nuevo León, 66238, México
pedro.valdes@udem.edu

[2] SUPA, School of Physics & Astronomy, Univeristy of St. Andrews, North Haugh, Fife, KY16 9SS, UK
fgr2@st-andrews.ac.uk

## ABSTRACT

Forty transits of the exoplanets HAT-P-12b, HAT-P-13b, HAT-P-16b, HAT-P-23b and WASP-10b were recorded with the 0.36m telescope at the Universidad de Monterrey Observatory. The images were captured with a standard Johnson-Cousins Rc and Ic and Sloan z' filters and processed to obtain individual light curves of the events. These light curves were successfully combined for each system to obtain a resulting one of higher quality, but with a slightly larger time sampling rate. A reduction by a factor of about four in per-point scatter was typically achieved, resulting in combined light curves with a scatter of ~1 mmag. The noise characteristics of the combined light curves were verified by comparing Allan variance plots of the residuals. The combined light curves for each system, along with radial velocity measurements from the literature when available, were modeled using a Monte Carlo method to obtain the essential parameters that characterize the systems. Our results for all these systems confirm the derived transit parameters (the planet-to-star radius ratio, $R_p/R_*$; the scaled semi-major axis, $a/R_*$; the orbital inclination, $i$; in some cases the eccentricity, $e$; and argument of periastron of the orbit, $\omega$), validating the methodology. This technique can be used by small college observatories equipped with modest-sized telescopes to help characterize known extrasolar planet systems. In some instances, the uncertainties of the essential transit parameters are also reduced. For HAT-P-23b, in particular, we derive a planet size $4.5 \pm 1.0\%$ smaller. We also derive improved linear periods for each system, useful for scheduling observations.

1. INTRODUCTION

The number of known exoplanets that transit in front of their stars continues to grow thanks to numerous ground-based and spacecraft search campaigns. For ground-based projects, the initial discovery announcement and characterization of a transiting extrasolar planet usually contains a few targeted follow-up observations from large aperture telescopes. However, further transit observations are desirable to independently confirm and refine the initial assessment of the system parameters derived from the light curve and radial velocity measurements. The number of transiting exoplanets known is large enough that almost every clear night there are observing opportunities available for any given observing site. This makes it worthwhile to organize a dedicated observing campaign of known extrasolar planet transits for the purpose of registering and analyzing light curves. One such effort is being carried out by the Extrasolar Transit Database (ETD; Poddany et al. 2010; http://var2.astro.cz/ETD/). At the Universidad de Monterrey (UDEM) Observatory we are also carrying out such an observing campaign. Our main goal is to observe multiple transits of the same target using standard photometric filters and to derive the important system parameters using our light curves combined with radial velocity measurements obtained from the literature.

In this paper, we analyze various transits of the extrasolar planet systems HAT-P-12, HAT-P-13, HAT-P-16, HAT-P-23, and WASP-10 in an attempt to improve the essential model parameters that characterize them. In Section 2, we discuss the methodology for the photometric observations and data reduction of the transits. In Section 3, we describe our method for combining several transit curves in order to obtain a curve with reduced noise, and we also derive an improved orbital period for the systems. In Section 4, we obtain the transit parameters from a best-fit models that also includes the radial velocity measurements. In Section 5, we compare our results with previous observations for each system, and in Section 6, we verify the validity of our light curve combination method using Allan variance plots and summarize our findings.

2. OBSERVATIONS AND DATA REDUCTION

All the transit light curves presented here were obtained using the UDEM Observatory telescope between 2008 and 2014. This is a small private college observatory (Minor Planet Center Code 720) located in the suburbs of Monterrey, México, at an altitude of 689 m. Most of the data were acquired using a standard Johnson-Cousins $Ic$-band filter (810 nm) on a 0.36 m reflector, with a 1280×1024 pixel CCD camera at a 1″.0 pixel$^{-1}$ scale resulting in a field of view of ~21′.3×17′.1. However, for HAT-P-23 we used an $Rc$-band filter (630 nm) because it was a dimmer target. In addition, two individual transits of HAT-P-13 and WASP-10 through a Sloan z'-band filter were also obtained. Redder bandpasses were elected for the observations as they provide increased photon counts in our detector and diminished turbulence effects. Longer-wavelength observations are also advantageous in that the target star exhibit decreased limb-darkening, which in turn provides a tighter constraint on the planet size. In general, the observations were slightly defocused to improve the photometric precision, and we always used on-axis guiding to maintain pointing stability. The exposure times varied between 30 and 90 s depending on the target star brightness and filter used, and the images were binned 2×2 to facilitate rapid readout (~3-4 s). Each observing session was programmed to accommodate the transit event and also to cover about one hour before ingress and one hour after egress. This was not always achievable due to weather and other constraints.

Standard dark current subtraction and twilight sky flat-field division process were performed on each image for calibration. Aperture photometry was carried out on the target star and 3-8 comparison stars of similar magnitude on every observing night. The measuring apertures used varied for each date due to the degree of defocus and weather conditions (seeing). Typical measuring aperture radii ranged between 3″.5 and 5″.5 (see Table 1 for individual values). For each night we reduced the data using various measuring apertures and number of comparison stars. We selected the best measuring aperture and comparison star combination to be the one that minimized the scatter of the resulting light curve. We also found that the best results were obtained by averaging the ratios of the target star to each selected comparison star. This produced smaller scatter than the standard method of ratioing the target star to the sum of all the comparison stars in the field, probably because we only used comparison stars that fell within a particular magnitude range (±~1.5 mag). Initially, we estimated the formal error for each photometric point as the standard deviation of the ratio to the individual comparison stars, divided by the square root of their number (error of the mean). However, visual inspection of the light curves clearly showed a larger variation between each contiguous measurement than the formal error indicated. Thus, we decided to quantify the scatter of the data not by the formal photometric error of the points but by the mean point-to-point difference of the data after removing a best-fit model (see below). This was a more conservative approach that best described the observed noise in the light curves. Scatter values varied depending mostly on the brightness of the star/filter combination, exposure time, and seeing conditions. Values ranged from 0.0030 (0.30%) in the best of conditions to typically between 0.0035 (0.35%) and 0.0045 (0.45%) in most cases.

After normalizing the target star to the comparison stars and averaging, some gradual variations as a function of time were often encountered. This is mainly caused by differential extinction between the transit and comparison stars, which generally have diverse spectral types. Bluer comparison stars are more affected by this atmospheric scattering effect than redder stars. Consequently, this variation was removed by using a linear air mass-dependent function of the form:

$$\Delta m = c(1 - X) \qquad (1)$$

where $\Delta m$ is the magnitude change applied to an individual measurement at air mass X, and $c$ is a best-fit constant necessary to remove the systematic effect and obtain a flat (horizontal) line for the out-of-transit baseline portions of the light curve. The best-fit model in equation (1) is determined from the out-of-transit data only. On some occasions, the trend observed on the out-of-transit baseline was better modeled by a simple linear fit that depended on the difference between the time of observation and the central meridian crossing of the star. These were associated with light-polluted skies, in particular, observing directions and weather conditions. We deemed this a more practical approach than attempting to photometrically correct for each comparison star spectral type difference based on their published filter magnitudes, for example, since these are often approximate and/or not available for fainter stars.

The relevant observing information for all dates is presented in Table 1.

## 3. LIGHT CURVE ANALYSIS

### 3.1 LIGHT CURVE COMBINATION

In principle, the final light curve for each date can be fitted independently with a transit model in order to obtain the parameters of the system. However, in practice, the amount of noise present would result in large uncertainties for the model parameters because of the modest telescope aperture used. Therefore, in order to decrease the noise in the light curve to be modeled and obtain better results, we decided to combine all available light curves for a given target. This is warranted because nearly all light curves obtained for a particular target have the same exposure times and similar number of comparison stars, which yielded similar scatter. However, in doing so, we were concerned that long-term brightness variations of the target stars might affect our resulting combined light curves, since deeper transits are measured for a given planet size when the host star is dimmer. This could be caused by starspot or other activity. Since we had differential photometry of the target stars with the same comparison stars over multiple epochs, we were able to record brightness variations of the host star or of a particular comparison star. The largest brightness variation of a target star was observed for HAT-P-16 and amounted to ~4%. Our observations were not frequent enough to detail if these brightness variations showed a trend or were periodic. However, these are within the observing noise limitations for individual transit depths and their effect is also diminished in the process of averaging multiple light curves. Combining all transit light curves into one should theoretically improve the scatter by a factor of $\sim\sqrt{n}$ and have a tendency to cancel out any systematic leftover trends that may be present after the reduction of the individual light curves.

In order to combine the light curves, we first need to co-register them by finding the mid-transit time for each event. In order to fit the observed transit light curves, we first created initial standard model light curves. These were constructed numerically as a tile-the-star procedure using the software package Binary Maker II (Bradstreet 2005). The initial system parameters used were obtained from the available literature and the linear limb-darkening function coefficient for this system was taken from Claret (2000). Small adjustments to the duration and depth of the model transits were necessary to optimize the fits and extract the best mid-transit time possible for each individual light curve. This was done by applying small (a few percent) multiplicative factors to both the depth and duration of the model transit. Precise model parameters were not required as this first approximation fit to the individual "noisy" light curves was only intended to derive the mid-transit time for the event. Best-fit models were obtained by minimizing the $\chi^2$ of the data. Again, to be conservative we used the average scatter of the photometry (mean point-to-point difference) to estimate the uncertainties because this value is larger than the formal error for each photometric measurement.

Once the initial mid-transit times for each system were obtained, a combined light curve was created by placing all the photometry points in a common time reference frame, centered on the mid-transit time for each event, and by averaging the data points falling within predetermined bins. This method generates a light curve more suitable for modeling since it contains fewer data points, exhibits less noise, and also maintains the general shape of the transit curve. However, care must be taken in selecting the bin size in order to simultaneously have sufficient amount of data in each bin, and also sufficient data points to define the critical ingress and egress portions of the light curve. Initially, we attempted to use bins defined with a fixed amount of time (we found out that two-minute bins best satisfied both criteria) but determined that this approach was unsuitable since some bins ended up containing more data points than others. This would result in the data points of the combined light curve to be modeled having unequal weights. Therefore, we decided to define bin sizes by the number of points to be averaged. For each system, we chose a fixed number of photometric points (ranging from 8 for HAT-P-12 to 19 for HAT-P-13) for the bins such that it would yield an average spacing of about two minutes between them. We then calculated the average flux value and average time from mid-transit for each bin to construct the combined light curve to be modeled. In obtaining these average values we weighed each point by the inverse of its individual uncertainty measurement.

Further refinements were performed to the individual light curve mid-transit times by subjecting the combined light curve to the same $\chi^2$ fit to the preliminary model and obtaining general multiplicative factors to both the depth and duration of the model transit for each system. These factors were then used on the individual light curves to derive improved mid-transit times and also generate a new combined light curve. This process was iterated until there was convergence between the individual mid-transit times and the best-fit multiplicative factors for the combined light curve. Table 1 also presents the observed transits and the final mid-transit times derived from fitting the models as explained above.

In all cases the resulting combined light curve exhibits a much reduced scatter compared with the individual observed light curves. Scatter values decreased by factors of 3.3-4.2 in the combined light curves, depending on the number of individual light curves combined, seeing conditions, magnitude limits, exposure times, etc. However, the value for the scatter of all of the combined light curves seems to level off at around 0.0010 (0.1%) for our system. This noise level seems to be close to the limit inherent in observing through an atmosphere as light curves produced with larger telescopes have difficulty improving on this value.

3.2 EPHEMERIS DETERMINATION

We derived an improved orbital period for the all the systems by performing a least-squares linear fit to our data and all other transit times available in the literature, weighing the individual mid-transit times by their uncertainties. See the individual discussions for each system in Section 5 for the references. When necessary we have converted the reported time frames (usually Barycentric Julian Date or Heliocentric Julian Date based on Coordinated Universal Time: BJD_UTC or HJD_UTC) to the improved Dynamical Time-based system (BJD_TDB) as suggested, for example, in Eastman et al. (2010). Our analysis for the studied systems yields the ephemeris shown in Table 2. In all instances the uncertainty is reduced. No clear transit-timing variations (TTVs) from a single period, which would suggest the presence of other planets in the systems, can be claimed at this time from this data. Other works have searched for an approximately constant period decrease (~0.01 s yr$^{-1}$) due to tidal effects for close orbiting planets, which appears as a gradual deviation from a linear ephemerides model usually fitted with a quadratic model (e.g., Adams et al. 2010). However, this requires observations over a large number of years and has been inconclusive for several objects due to uncertainties in transit timings (e.g., Ricci et al. 2015)

## 4. TRANSIT MODELING

To estimate the essential transit parameters (the planet-to-star radius ratio, $R_p/R_*$; the scaled semimajor axis, $a/R_*$; and the orbital inclination, $i$) from the combined light curves of each system, we use the software package EXOFAST designed for IDL (Eastman et al. 2013). It uses the now standard Mandel & Agol (2002) models for the transit light curve calculation and implements a differential evolution Markov Chain Monte Carlo algorithm to find the best-fit parameters for the observed light curve. This analysis evaluates different combinations of parameters until it converges into an optimum solution. Such a scheme also allows estimating the uncertainty of the transit parameters.

To start the analysis we adopted the spectroscopic parameters (stellar effective temperature, $T_{eff*}$; surface gravity, $\log g_*$; and metallicity, [Fe/H]) and published uncertainties for each system usually reported in the discovery paper, and also used quadratic limb-darkening coefficients for the stars from Claret (2000). Southworth (2008) concluded that there is no significant difference in using either a single linear (u) or two quadratic (a and b) coefficients to describe the stellar limb-darkening in the analysis of high-quality ground-based data. These model constants are presented in Table 3 for each system. We also adopted our derived periods from Section 3.

Initially, we also adopted the orbital parameters (namely, the eccentricity, $e$, and argument of periastron, $\omega$, of the star´s orbit) from the original discovery paper and ran models with these values fixed. However, EXOFAST also has the option of incorporating radial velocity measurements and deriving self-consistent solutions for the stellar, transit, and radial velocity parameters. When possible, we have incorporated the radial velocities to better characterize the extrasolar transiting planet systems. See the discussions in Section 5 for each individual case in which we performed this complete analysis.

## 5. RESULTS AND DISCUSSION

### 5.1 HAT-P-12

We registered six transits for HAT-P-12 through a standard Johnson-Cousins Ic filter between 2011 and 2014 (see Table 1 and Figure 1a) that we used to derive the combined light curve shown at the bottom of Figure 1a. Average point-to-point variation values ranged between 0.0031 and 0.0045 (0.31% - 0.45%) for the individual light curves, while the value for the combined one was 0.0012 (0.12%), for an improvement of a factor of 3.3. Our observations also show that the host star did not vary in brightness with respect to the comparison stars when out of transit, which lends confidence to combining the individual light curves. For the calculation of the period and epoch we used, aside from our six observations, four mid-transit times of Hartman et al. (2009; as presented by Lee et al. 2012), three from Lee et al. (2012) and one from Sada et al. (2012). Our result (P = 3.2130589 ± 0.0000003 days and $T_c$ = 2454419.19584 ± 0.00009 BJD_TDB) agrees with the one obtained by Lee et al. (2012) and with the one by Todorov et al. (2013), both of which also included reanalyzed amateur light curves from the ETD website. No long-term TTVs are evident at this time (see Figure 1b) and there are few mid-transit times available for the determination of short-term variations.

Hartman et al. (2009) derive an orbit with zero eccentricity from the radial velocity data presented in the discovery paper. Recently, Knutson et al. (2014) include a few more radial velocity measurements, but they also calculate an eccentricity for the orbit that is consistent with zero. Thus, we also adopt a circular orbit for this system in our modeling. Our results (see Table 4) are in general agreement with those obtained by Hartman et al. (2009), Lee et al. (2012), and Sada et al. (2012). However, the recent analysis of a near-IR transmission spectrum from Hubble Space Telescope data by Line et al. (2013) yields a planet-to-star radius ratio that is ~2.5% smaller than the others.

5.2 HAT-P-13

We observed nine transits of this system between 2010 and 2014 (see Table 1 and Figures 2a and 2b). Average point-to-point variation values ranged between 0.0032 and 0.0044 (0.32% – 0.44%) for the individual light curves, while the value for the combined one was 0.0010 (0.10%), for an improvement of a factor of 3.9. Eight of these observations were recorded through a standard Johnson-Cousins Ic filter and another one through a Sloan z' filter (not shown). This last one was used only to help define a linear period for the system and not to derive a combined light curve. Besides the transiting planet, radial velocity measurements of this system have revealed the presence of a second planet in the system that is not known to transit the disk of the star (Bakos et al. 2009) and perhaps a third one (Winn et al. 2010, Knutson et al. 2014) as well. So far the gravitational influence of this second planet has not been definitely observed as mid-transit-timing variations on the transiting planet. Initially, there were claims of possible statistically significant TTVs (Nascimbeni et al. 2011; Pál et al. 2011), but as the number of registered transits grew, this claim started to be in doubt (Fulton et al. 2011; Southworth et al. 2012). Aside from the issue of analyzing a small number of mid-transit times to find possible statistically significant deviations from a linear period, the transits of HAT-P-13b, in particular, are fairly shallow (~0.75%) and have long ingress and egress times (~30 minutes). This results in relatively large uncertainties (±~60 s in the best of circumstances, and usually 50% larger) in the mid-transit time determinations from ground-based light curve observations. Payne & Ford (2011), for example, suggest that mid-transit timings would need to have uncertainties at least four to six times smaller than current observations to confidently use the TTVs to help constrain the orbit of the perturbing planet. This seems hard to achieve from ground-based observations. Because of this, in this work we have chosen to fit a linear period and epoch to the available data. Our observations also extend the timeline of recorded transits to six years and reduce the uncertainties compared with previous works. The result is of use to predict future transits of HAT-P-13b. Besides our nine mid-transit times (see Table 1), we have also used 31 others available from light curves published in Bakos et al. (2009; as presented in Pál et al. 2011), Szabó et al. (2011), Nascimbeni et al. (2011), Pál et al. (2011), Fulton et al. (2011), and Southworth et al. (2012). Our derived linear period and epoch (P = 2.9162433 ± 0.0000012 days & $T_c$ = 2455176.53893 ± 0.00022 BJD_TDB) (see Figure 2c) is in general agreement with the latest determination by Southworth et al. (2012).

In our combined light curve modeling for this system we adopted the eccentricity, $e$, and angle of periastron, $\omega$, from the calculations of Winn et al. (2010) and Knutson et al. (2014) because the software cannot derive these values from radial velocity measurements in which there are noticeable gravitational effects from another planet. Our results (see Table 4) are similar in value and uncertainties to those derived originally in Bakos et al. (2009) mainly because we adopted their spectroscopically derived values for $T_{eff*}$, log $g_*$, and [Fe/H].

5.3 HAT-P-16

We recorded four transits of the HAT-P-16 system between 2010 and 2013 using a standard Ic filter (see Table 1 and Figure 3a). Average point-to-point variation values ranged between 0.0030 and 0.0042 (0.30% – 0.42%) for the individual light curves, while the value for the combined one was 0.0010 (0.10%), for an improvement of a factor of 3.6. Our derived linear period and epoch (P = 2.7759704 ± 0.0000007 and $T_c$ = 5027.59292 ± 0.00019) also used mid-transit times from Buchhave et al. (2010) and Ciceri et al. (2013). The amateur mid-transit times used by Ciceri et al. (2013) were not included in this study. The result is consistent with previous estimates and reduces the uncertainties further because of the increased time coverage. There are too few observations available for this system to definitively detect a TTV signal. No clear sign of one is evident from Figure 3b.

In modeling our combined light curve we also used radial velocity information from Buchhave et al. (2010), Moutou et al. (2011) and Knutson et al. (2014) to solve for the eccentricity, $e$, and argument of periastron, $\omega$. The results shown in Table 4 for HAT-P-16 clearly confirm the results of Buchhave et al. (2010).

5.4 HAT-P-23

We obtained 11 transits for HAT-P-23 through a standard Rc filter (see Table 1 and Figures 4a and 4b) from observations between 2011 and 2014 and derived a resulting light curve which, upon analysis, yields results in line with those obtained with combining only the first four light curves as presented in Ramón-Fox & Sada (2013). Average point-to-point variation values ranged between 0.0036 and 0.0053 (0.36% – 0.53%) for the individual light curves, while the value for the combined one was 0.0010 (0.10%), for an improvement of a factor of 4.2. The period and epoch (P = 1.2128867 ± 0.0000002 days and $T_c$ = 4852.26548 ± 0.00017 BJD_TDB) were derived with our 11 mid-transit times (including slight mid-transit times revision for the four from Ramón-Fox & Sada, 2013) and the ephemeris given in the original discovery paper by Bakos et al. (2011). No other transits were found in the refereed literature. A six-year baseline allows us to constrain the uncertainty further, and no deviations from a single linear period are evident from the available data (see Figure 4c).

Unlike in the analysis by Ramón-Fox & Sada (2013), where only a combined light curve from four individual transits was modeled and the orbital parameters $e$ and $\omega$ were fixed, in the present work we also included the available radial velocity information in order to constrain the model parameters. These were taken from Bakos et al. (2011) and from Moutou et al. (2011), who determined the system´s spin-orbit inclination. The simultaneous analysis of the combined light curve and radial velocity information for the system allowed for a more realistic estimation of the uncertainties for our resulting parameters (see Table 4) than in Ramón-Fox & Sada (2013). Our inclination and scaled semi-major axis of the system agree with those of Bakos et al. (2011) and Ramón-Fox & Sada (2013), and the eccentricity and argument of periastron also agree with Bakos et al. (2011) with reduced uncertainties due to the inclusion of the Moutou et al. (2011) radial velocity data set.

We do note a difference in the planet-to-star radius ratio $R_p/R_*$ for the system. Our value is about $4.5 \pm 1.0\%$ lower than the one initially derived in Bakos et al. (2011), and in general agreement with Ramón-Fox & Sada (2013). Through our observing campaign we noticed that the star in this system varied in brightness by ~ 3%, probably due to starspot activity, and suspected that perhaps our individual light curves would have different transit depths due to this. However, our analysis to find a correlation between individual transit depths and stellar brightness variations yielded null results. Thus, we conclude that the combination of individual transit data was a valid attempt at deriving a typical light curve for the transiting system that can be analyzed to obtain the main parameters for the system.

Our smaller planet-to-star radius ratio is in line with the expectation of Fortney et al. (2008), which predicted a smaller planet size (~8.4% smaller than observed by Bakos et al., 2011) from their theoretical models for a planet of this mass. However, we could not improve on the uncertainty regarding the inclination of the system, which would have helped answer the question regarding the projected angle between the orbital plane and the stellar equatorial plane outlined in Moutou et al. (2011).

Recently, O´Rourke et al. (2014) performed near-IR secondary eclipse photometry of HAT-P-23 and concluded from their mid-eclipse time observations that this system likely has a circular orbit. Our decreased uncertainty for the orbital value of $e$ still advocates for non-circularity, although more radial velocity information is still desirable to settle the issue.

5.5 WASP-10

We observed nine transits of WASP-10 through a standard Ic filter (see Table 1 and Figures 5a and 5b) between 2008 and 2014. Average point-to-point variation values ranged between 0.0031 and 0.0047 (0.31% – 0.47%) for the individual light curves, while the value for the combined one was 0.0010 (0.10%), for an improvement of a factor of 3.4. Another transit observed through a z' filter (2009 September 16 – not shown) was also obtained but was not used in the construction of the combined light curve. However, it was used to help derive a period for the system. Our result (P = 3.0927295 ± 0.0000003 days and $T_c$ = 4664.03804 ± 0.00006 BJD_TDB) was derived from our 10 light curves plus 30 others obtained from Christian et al. (2009), Johnson et al. (2009), Dittmann et al. (2010), Maciejewski et al. (2011a; which included four modified measurements from Krejčová et al. 2010), Maciejewski et al. (2011b), Sada et al. (2012), and Barros et al. (2013). Maciejewski et al. (2011a, 2011b) in particular, perform transit-timing analysis of the data and found small periodic variations presumably due to another planet in the system. However, Barros et al. (2013), upon thorough reanalysis of the data, do not find such variations and attribute them to starspot occultation features, especially during ingress and/or egress, or to systematics. We provide 10 other mid-transit times that extend the observational baseline to 7 years and reduces the uncertainty in the linear period fit.

This system exhibits a very deep transit (~3%) with sharp and well-defined short ingress and egress times (~20 minutes), and it is located on a field with several comparison stars of similar magnitude. This makes it particularly well suited for observation and modeling. This has been done by not only Christian et al. (2009), but also by Johnson et al. (2009), Dittmann et al. (2010), Krejčová et al. (2010), Maciejewski et al. (2011b) and Barros et al. (2013). Our results, in particular, match those of Johnson et al. (2009), who analyzed a single high signal-to-noise light curve. For our modeling, we adopted the eccentricity, $e$, and angle of periastron, $\omega$, values from Knutson et al. (2014) since they discovered a radial velocity trend in the data that suggests the presence of another planet and the software is not designed to handle such a circumstance. However, it must be said that Husnoo et al. (2012) advocate for a circular orbit for this system, and the recent observation of a secondary eclipse for WASP-10 by Cruz et al. (2014) was insufficient to differentiate between a circular and an eccentric orbit.

The star in this system was initially suspected (Dittmann et al. 2009) and later confirmed (Smith et al. 2009; Maciejewski et al. 2011a, 2011b; Barros et al. 2013) to exhibit starspot activity that affects not only the radial velocity measurements and mid-transit timings, but also the modeling of the system parameters. Our light curves show no clear evidence of starspot activity to the noise level of the individual light curves and we detected less than ~2% intensity variation of the star between our observations. Thus, we felt confident that our combined light curve yielded modeling results that are in line with all other work.

## 6. ALLAN VARIANCE, SUMMARY, AND CONCLUSIONS

We have shown that combining light curves of extrasolar planet transits obtained with modest equipment, like that found in small college observatories located in suburban areas, results in light curves with improved signal-to-noise levels that are suitable for model analysis. However, the observations and data reduction phases of the process have to be as consistent and systematic as possible to yield useful combined light curves. In particular, care must be taken in selecting comparison stars of similar magnitude as the target, apertures in the photometric measurements, and detrending due to differential extinction and light pollution needs to be made in most circumstances.

To test our methodology, we have combined a total of 38 light curves for the exoplanet systems HAT-P-12 (6, Ic), HAT-P-13 (8, Ic), HAT-P-16 (4, Ic), HAT-P-23 (11, Rc), and WASP-10 (9, Ic) obtained at the UDEM Observatory (MPC 720) between 2009 and 2014 (see Table 1). We have also used two additional light curves taken with a Sloan z' filter (one for HAT-P-13 and one for WASP-10) to help derive, with the aid of the mid-transit times presented in the literature, the linear period and epochs for the transiting exoplanets. These results for determining periods and epochs (see Table 2) match those of the literature and reduce the uncertainties further due to the longer timeline of observations.

In order to further explore the notion that the combination of individual light curves does tend to cancel out systematic errors present in the original data we have created Allan variance plots of the residuals (observed minus model) corresponding to each of the individual light curves and the residuals corresponding to the resulting combined light curves. For this, we followed the formalism detailed in Carter et al. (2009). We have used their definition of Allan variance as stated in their Equation (8) to plot its value as a function of lag. This is shown in Figure 6. The dotted lines represent all the individual light curve residuals and the labeled solid lines are the combined light curve residuals. The Allan variance values of the residuals of the original light curves as a function of lag exhibit white (uncorrelated) noise characteristics. We interpret this as a sign that the detrending due to differential extinction used is mostly successful at eliminating tendencies in the original light curves. The Allan values of the residuals of the combined light curves also exhibit the same white noise characteristics at low and high lag values with some flicker noise being present at mid-lag values in some cases. This is particularly noticeable for the HAT-P-23 curve. We are unsure as of its source but theorize that it may be related to reaching the limitations of the method as this is the resulting light curve with the largest number of individual light curves combined. There is also about an order of magnitude difference throughout in the Allan variance values of the original light curve residuals compared with the resulting light curve residuals, indicating that the noise level has decreased substantially due to the combination of the individual light curves.

For modeling the light curves, we have used the software package EXOFAST (Eastman et al. 2013) and have set the spectroscopic parameters ($T_{eff*}$, log $g_*$ and [Fe/H]) and quadratic limb-darkening coefficients (see Table 3). When possible in the modeling process, we have also included the radial velocity information available in the literature, although at times the eccentricity, $e$, and argument of periastron, $\omega$, were also adopted from the literature. In all cases, our results (see Table 4) agree very well with those obtained in previous works that used observations from larger telescopes and/or better observing sites. In some cases, we obtain smaller uncertainties to these values. These results validate our methodology. We also find that after combining 6-8 individual light curves the uncertainties in the resulting model parameters are improved only marginally. This is a combined effect of the signal to noise of the final light curve being sensitive to the square root of the number of observations and also the inherit limitation of observing through an atmosphere.

In our analysis, we found significant differences for only one system. For HAT-P-23, we derive a planet-to-star radius ratio, $R_p/R_*$, which is 4.5±1.0% lower than the one available in the literature and reduced the uncertainties in values for the orbital eccentricity, $e$, and argument of periastron, $\omega$.

ACKNOWLEDGMENTS

F.G.R.F. acknowledges financial support from the ERC grant ECOGAL and the University of St. Andrews.

# REFERENCES

- Adams, E. R., Lopez-Morales, M., Elliot, J. L., Seager, S., & Osio, D. J. 2010, ApJ, 721, 1829.
- Bakos, G. Á., Howard, A. W., Noyes, R. W., Hartman, J., Torres, G., Kovács, G., Fischer, D. A., Latham, D. W., Johnson, J. A., Marcy, G. W., Sasselov, D. D., Stefanik, R. P., Sipõcz, B., Kovács, G., Esquerdo, G. A., Pál, A., Lázár, J., Papp, I., and Sári, P. 2009, ApJ, 707, 446.
- Bakos, G. Á., Hartman, J., Torres, G., Latham, D. W., Kovács, G., Noyes, R. W., Fischer, D. A., Johnson, J. A., Marcy, G. W., Howard, A. W., Kipping, D., Esquerdo, G. A., Shporer A., Béky, B., Buchhave, L. A., Perumpilly, G., Everett, M., Sasselov, D. D., Stefanik, R. P., Lázár, J., Papp, I. & Sári, P. 2011, ApJ, 742, 116.
- Barros, S. C. C., Boué, G., Gibson, N. P., Pollacco, D. L., Santerne, A., Keenan, F. P., Skilllen, I. and Street, R. A. 2013, MNRAS, 430, 3032.
- Bradstreet, D. H. 2005, SASS, 24, 23.
- Buchhave, L. A., Bakos, G. Á., Hartman, J. D., Torres, G., Kovács, G., Latham, D. W., Noyes, R. W., Esquerdo, G. A., Everett, M., Howard, A. W., Marcy, G. W., Fischer, D. A., Johnson, J. A., Andersen, J., Fũrész, G., Perumpilly, G., Sasselov, D. D., Stefanik, R. P., Béky, B., Lázár, J., Papp, I. and Sári, P. 2010, ApJ, 720, 1118.
- Carter, J. A., Winn, J. N., Gilliland, R., and Holman, M. J. 2009, ApJ, 696, 241.
- Christian, D. J., Gibson, N. P., Simpson, E. K., Street, R. A., Skillen, I., Pollacco, D., Collier-Cameron, A., Joshi, Y. C., Keenan, F. P., Stempels, H. C., Haswell, C. A., Horne, K., Anderson, D. R., Bentley, S., Bouchy, F., Clarkson, W. I., Enoch, B., Henn, L., Hébrard, G., Hellier, C., Irwin, J., Kane, S. R., Lister, T. A., Loeillet, B., Maxted, P., Mayor, M., McDonald, I., Moutou, C., Norton, A. J., Parley, N., Pont, F., Queloz, D., Ryans, R., Smalley, B., Smith, A. M. S., Todd, I., Udry, S., West, R. G., Wheatley, P. J., and Wilson, D. M. 2009, MNRAS, 392, 1585.
- Ciceri, S., Mancini, L., Southworth, J., Nikolov, N., Bozza, V., Bruni, I., Calchi Novati, S., DÁgo, G., and Henning, Th. 2013, A&A, 557, A30.
- Claret, A. 2000, A&A 363, 1081.
- Cruz, P., Barrado, D., Lillo-Box, J., Diaz, M., Birkby, J., López-Morales, M., Hodgkin, S. and Fortney, J. J. 2014, arXiv: 1412:2996v1.
- Dittmann, J. A., Close, L. M., Scuderi, L. J., and Morris, M. D. 2010, ApJ, 717, 235.
- Eastman, J., Gaudi, B. S., & Agol. E. 2013, PASP, 125, 83.
- Eastman, J., Siverd, R. & Gaudi, B. S. 2010, PASP, 122, 935.
- Fortney, J. J., Lodders, K., Marley, M. S., & Freedman, R. S. 2008, ApJ, 678, 1419.
- Fulton, B. J., Shporer, A., Winn, J. N., Holman, M. J., Pál, A., and Gazak, J. Z. 2011, ApJ, 142, 84.
- Hartman, J. D., Bakos, G. Á., Torres, G., Kovács, G., Noyes, R. W., Pál, A., Latham, D. W., Sipõcz, B., Fischer, D. A., Johnson, J. A., Marcy, G. W., Butler, R. P., Howerd, A. W., Esquerdo, G. A., Sasselov, D. D., Kovács, G., Stefanik, R. P., Fernandez, J. M., Lázár, J., Papp, I. and Sári, P. 2009, ApJ, 706, 785.

- Husnoo, N., Pont, F., Mazeh, T., Fabrycky, D., Hébrard, G., Bouchy, F. and Shporer, A. 2012, MNRAS, 422, 3151.
- Johnson, J. A., Winn, J. N., Cabrera, N. E., and Carter, J. A. 2009, ApJ, 692, L100.
- Knutson, H. A., Fulton, B. J., Montet, B. T., Kao, M., Ngo, H., Howard, A. W., Crepp, J. R., Hinkley, S., Bakos, G. Á., Batygin, K., Johnson, J. A., Morton, T. D., and Muirhead, P. S. 2014, ApJ, 785, 126.
- Krejčová, T., Budaj, J, and Krushevska, V. 2010, Contrib. Astron. Obs. Skalnaté Pleso 40, 77.
- Lee, J. W., Youn, J.-H., Kim, S.-L., Lee, C.-U. and Hinse, T. C. 2012, AJ 143, 95.
- Line, M. R., Knutson, H., Deming, D., Wilkins, A. and Desert, J.-M. 2013, ApJ, 779, 183.
- Maciejewski, G., Dimitrov, D., Neuhäuser, R., Tetzlaff, N., Niedzielski, A., Raetz, St., Chen, W. P., Walter, F., Marka, C., Baar, S., Krejcová, T., Budaj, J., Krushevska, V., Tachihara, K., Takahashi, H., and Mugrauer, M. 2011a, MNRAS, 411, 1204.
- Maciejewski, G., Raetz, St., Nettelmann, N., Seeliger, M., Adam, Ch., Nowak, G. and Neuhäuser, R. 2011b, A&A, 535, A7.
- Mandel, K. & Agol, E. 2002, ApJ, 580, L171.
- Moutou, C., Díaz, R. F., Udry, S., Hébrard, G., Bouchy, F., Santerne, A., Ehrenreich, D., Arnold, L., Boisse, I., Bonfils, X., Delfosse, X., Eggenberger, A., Forveille, T., Lagrange, A.-M., Lovis, C., Martinez, P., Pepe, F., Perrier, C., Queloz, D., Santos, N. C., Ségransan, D. 2011, A&A, 533, A113.
- Nascimbeni, V., Piotto, G., Bedin, L. R., Damasso, M., Malavota, L., and Borsato, L. 2011, A&A, 532, A24.
- O'Rourke, J. G., Knutson, H. A., Zhao, M., Fortney, J. J., Burrows, A., Agol, E., Deming, D., Désert, J.-M., Howard, A. W., Lewis, N. K., Showman, A. P., and Todorov, K. O. 2014, ApJ, 781, 109.
- Pál, A., Sárneczky, K., Szabó, G. M., Szing, A., Kiss, L. L., Mező, G., and Regály, Z. 2011, MNRAS, 413, L13.
- Payne, M. J., and Ford, E. B. 2011, ApJ, 729, 98.
- Poddany, S., Brat, L. & Pejcha, O. 2010, New Astronomy, 15, 297.
- Ramón-Fox, F. G. & Sada, P. V. 2013, RMAA, 49, 71.
- Ricci, D., Ramón-Fox, F. G., Ayala-Loera, C., Michel, R., Navarro-Meza, S., Fox-Machado, L. & Reyes-Ruiz, M. 2015, PASP, 127, 143
- Sada, P. V., Deming, D., Jennings, D. E., Jackson, B., Hamilton, C. M., Fraine, J., Peterson, S. E., Haase, F., Bays, K., Lunsford, A., O´Gorman, E. 2012, PASP, 124, 212.
- Smith, A. M. S., Hebb, L., Collier Cameron, A., Anderson, D. R., Lister, T. A., Hellier, C., Pollacco, D., Queloz, D., Skillen, I., and West, R. G. 2009, MNRAS, 398, 1827.
- Southworth, J. 2008, MNRAS, 386, 1644.
- Southworth, J., Bruni, I., Mancini, L., and Gregorio, J., 2012, MNRAS, 420, 2580.
- Szabó, Gy. M., Kiss, L. L., Benkő, J. M., Mező, Gy., Nuspl, J., Regály, Zs., Sárneczky, K., Simon, A. E., Leto, G., Zanmar Sanchez, R., Ngeow, C. -C., Kővári, Zs., and Szábo, R. 2010, A&A, 523, A84.

- Todorov, K. O., Deming, D., Knutson, H. A., Burrows, A., Fortney, J. J., Kewis, N. K., Cowan, N. B., Agol. E., Desert, J.-M., Sada, P. V., Charbonneau, D., Laughlin, G., Langton, J. and Showman, A. P. 2013, ApJ, 770, 102.
- Winn, J., N., Johnson, J. A., Howard, A. W., Marcy, G. W., Bakos, G. Á., Hartman, J., Torres, G., Albrecht, S., and Narita, N. 2010, ApJ, 718, 575.

TABLE 1

UDEM OBSERVATORY EXTRASOLAR PLANET TRANSITS

| Name | Date | Filter | Exp. (s) | N | Cp. St. | M.Ap.R. (″) | Base. Fit | Tc - 2450000.0 (BJD_TDB) | Notes |
|---|---|---|---|---|---|---|---|---|---|
| HAT-P-12 | 2011 Apr 27 | Ic | 90 | 170 | 7 | 4.4 | None | 5678.71462 ± 0.00047 | |
| | 2012 Apr 24 | Ic | 90 | 165 | 5 | 4.7 | X, c = -0.0045 | 6041.79018 ± 0.00037 | |
| | 2012 May 23 | Ic | 90 | 173 | 7 | 4.9 | L, m = -0.0100 | 6070.70792 ± 0.00035 | |
| | 2012 Jun 08 | Ic | 90 | 172 | 8 | 5.6 | X, c = +0.0009 | 6086.77329 ± 0.00051 | |
| | 2013 Jun 19 | Ic | 90 | 155 | 7 | 5.0 | X, c = +0.0028 | 6462.70081 ± 0.00050 | |
| | 2014 Jun 01 | Ic | 90 | 173 | 7 | 4.3 | X, c = -0.0026 | 6809.71238 ± 0.00045 | |
| HAT-P-13 | 2010 Feb 13 | z' | 60 | 290 | 1 | 5.9 | None | 5240.69554 ± 0.00197 | Period only |
| | 2011 Jan 23 | Ic | 45 | 389 | 3 | 5.3 | L, m = -0.0270 | 5584.81245 ± 0.00118 | |
| | 2011 Jan 26 | Ic | 45 | 374 | 6 | 5.8 | None | 5587.73154 ± 0.00151 | |
| | 2011 Mar 02 | Ic | 30 | 542 | 6 | 4.2 | X, c = +0.0040 | 5622.72289 ± 0.00120 | |
| | 2012 Jan 08 | Ic | 45 | 335 | 7 | 5.4 | X, c = -0.0042 | 5934.76202 ± 0.00155 | |
| | 2013 Jan 24 | Ic | 45 | 417 | 7 | 4.4 | L, m = -0.0105 | 6316.79247 ± 0.00117 | |
| | 2013 Feb 28 | Ic | 45 | 227 | 6 | 4.9 | None | 6351.78519 ± 0.00211 | Ingress only |
| | 2013 Mar 03 | Ic | 45 | 408 | 7 | 5.1 | L, m = -0.0048 | 6354.70220 ± 0.00112 | |
| | 2014 Feb 13 | Ic | 45 | 295 | 3 | 4.3 | L, m = +0.0100 | 6701.73623 ± 0.00139 | |
| HAT-P-16 | 2010 Oct 02 | Ic | 40 | 414 | 6 | 4.9 | X, c = -0.0032 | 5471.74748 ± 0.00047 | |
| | 2011 Nov 28 | Ic | 40 | 412 | 8 | 4.9 | X, c = +0.0031 | 5893.69673 ± 0.00065 | |
| | 2012 Sep 20 | Ic | 40 | 416 | 8 | 5.2 | X, c = -0.0036 | 6190.72516 ± 0.00059 | |
| | 2013 Oct 08 | Ic | 40 | 439 | 8 | 5.4 | L, m = +0.0530 | 6573.80636 ± 0.00057 | |
| HAT-P-23 | 2011 Jun 03 | Rc | 60 | 223 | 8 | 4.0 | L, m = -0.0018 | 5715.84175 ± 0.00060 | |
| | 2011 Aug 04 | Rc | 60 | 238 | 8 | 4.2 | X, c = +0.0047 | 5777.69855 ± 0.00058 | |
| | 2011 Aug 16 | Rc | 60 | 220 | 7 | 3.2 | X, c = -0.0026 | 5789.82583 ± 0.00062 | |
| | 2011 Aug 21 | Rc | 60 | 210 | 7 | 3.7 | X, c = +0.0106 | 5794.67924 ± 0.00054 | |

| Name | Date | Filter | Exp. | N | Cp. St. | M.Ap.R. | Base. Fit | Tc - 2450000.0 | Notes |
|---|---|---|---|---|---|---|---|---|---|
| | 2012 Sep 06 | Rc | 60 | 239 | 8 | 3.9 | X, c = -0.0074 | 6176.73801 ± 0.00064 | |
| | 2012 Sep 23 | Rc | 60 | 268 | 8 | 5.3 | X, c = -0.0089 | 6193.71578 ± 0.00075 | Scatter |
| | 2012 Oct 04 | Rc | 60 | 223 | 8 | 4.5 | X, c = -0.0063 | 6204.63395 ± 0.00056 | |
| | 2012 Oct 10 | Rc | 60 | 134 | 8 | 4.5 | None | 6210.70212 ± 0.00093 | Ingress only |
| | 2013 Aug 08 | Rc | 60 | 208 | 7 | 4.1 | X, c = -0.0124 | 6512.70668 ± 0.00058 | Gaps |
| | 2014 Aug 25 | Rc | 60 | 236 | 8 | 4.4 | X, c = +0.0007 | 6894.76653 ± 0.00052 | |
| | 2014 Sep 11 | Rc | 60 | 236 | 8 | 5.0 | X, c = +0.0012 | 6911.74741 ± 0.00060 | |
| WASP-10 | 2008 Aug 10 | Ic | 60 | 263 | 5 | 3.9 | X, c = +0.0059 | 4688.77968 ± 0.00040 | |
| | 2008 Nov 17 | Ic | 60 | 219 | 5 | 3.5 | X, c = +0.0056 | 4787.74815 ± 0.00033 | No egress base |
| | 2009 Sep 16 | z' | 60 | 249 | 5 | 4.1 | X, c = -0.0070 | 5090.83487 ± 0.00052 | Period only |
| | 2010 Aug 13 | Ic | 90 | 160 | 6 | 4.9 | X, c = +0.0038 | 5421.75689 ± 0.00035 | |
| | 2010 Sep 16 | Ic | 90 | 159 | 7 | 5.0 | L, m = +0.0140 | 5455.77761 ± 0.00031 | |
| | 2010 Oct 17 | Ic | 90 | 181 | 8 | 5.6 | L, m = +0.0250 | 5486.70317 ± 0.00035 | |
| | 2011 Oct 20 | Ic | 90 | 166 | 8 | 5.5 | X, c = -0.0025 | 5854.73850 ± 0.00031 | |
| | 2012 Sep 18 | Ic | 90 | 165 | 8 | 5.2 | L, m = +0.0040 | 6188.75376 ± 0.00032 | |
| | 2012 Oct 22 | Ic | 90 | 137 | 8 | 4.1 | X, c = -0.0006 | 6222.77315 ± 0.00031 | No egress base |
| | 2014 Nov 25 | Ic | 90 | 150 | 8 | 5.0 | X, c = -0.0055 | 6986.67714 ± 0.00040 | |

| | |
|---|---|
| Name: | Name of the extrasolar planet system. |
| Date: | Observation date (Universal Time). |
| Filter: | Standard Photometric filter used for the observations. |
| Exp.: | Exposure time of each image in seconds. |
| N: | Number of individual frames acquired and measured during the observation. |
| Cp. St.: | Number of comparison stars used to obtain the light curve of the target star. |
| M.Ap.R.: | Measuring aperture radius (in arcseconds) used for all the comparison and target stars in all the frames. |
| Base. Fit: | Type of fit used on the out-of-transit baseline. |
| | X = air mass dependence as per Equation (1) with the constant $c$ stated. |
| | L = linear fit with meridian crossing as pivot point and slope $m$ stated. |
| Tc - 2450000.0: | Barycentric Julian Date based on Dynamical Time for the mid-transit time of the system. |
| Notes: | Period only = Used to obtain period of system but not combined light curve. |

Ingress only = No egress available due to clouds.
Scatter = Larger scatter than usual due to weather conditions.
Gaps = Small gaps in the middle of the light curve due to passing clouds.
No egress base = Observations interrupted just after egress due to clouds.

TABLE 2

DERIVED SYSTEM PERIODS

| Name | Number of Data Sets | Time Coverage (years) | Period (days) | Tc - 2450000.0 (BJD_TDB) |
|---|---|---|---|---|
| HAT-P-12 | 14 | 8 | 3.2130589(3) | 4419.19584(9) |
| HAT-P-13 | 40 | 6 | 2.9162433(12) | 5176.53893(22) |
| HAT-P-16 | 11 | 5 | 2.7759704(7) | 5027.59292(19) |
| HAT-P-23 | 12 | 6 | 1.2128867(2) | 4852.26548(17) |
| WASP-10 | 40 | 7 | 3.0927295(3) | 4664.03804(6) |

TABLE 3

FIXED MODEL PARAMETERS

| Name | $T_{eff*}$ (K) | $\log_{g*}$ | [Fe/H] | Limb Darkening (1) | | |
|---|---|---|---|---|---|---|
| | | | | u | a | b |
| HAT-P-12 (2) | 4650± 60 | 4.61±0.03 | -0.29±0.05 | 0.593 | 0.390±0.006 | 0.239±0.006 |
| HAT-P-13 (3) | 5653± 90 | 4.13±0.04 | 0.41±0.08 | 0.506 | 0.271±0.006 | 0.321±0.010 |
| HAT-P-16 (4) | 6158± 80 | 4.34±0.03 | 0.17±0.08 | 0.492 | 0.193±0.005 | 0.353±0.006 |
| HAT-P-23 (5) | 5905± 80 | 4.33±0.06 | 0.15±0.04 | 0.610 | 0.324±0.001 | 0.339±0.001 |
| WASP-10 (6) | 4675±100 | 4.63±0.01 | 0.03±0.20 | 0.560 | 0.404±0.005 | 0.232±0.005 |

(1) Interpolated from the data provided in Claret 2000.
(2) Hartman et al. 2009
(3) Bakos et al. 2009
(4) Buchhave et al. 2010
(5) Bakos et al. 2011.
(6) As reported by Johnson et al. 2009

TABLE 4

MODELED MAIN SYSTEM PARAMETERS

| Name | $i$ (degrees) | $a/R_*$ | $R_p/R_*$ | $e$ | $\omega$ (degrees) | |
|---|---|---|---|---|---|---|
| HAT-P-12 | $88.17^{+0.34}_{-0.28}$ | $12.00^{+0.36}_{-0.33}$ | $0.1400^{+0.0012}_{-0.0012}$ | Assumed Circular Orbit | | (1) |
| HAT-P-13 | $82.81^{+0.60}_{-0.59}$ | $5.61^{+0.23}_{-0.22}$ | $0.0860^{+0.0012}_{-0.0012}$ | $0.0133 \pm 0.0045$ | $197^{+32}_{-37}$ | (2) |
| HAT-P-16 | $88.03^{+1.00}_{-0.74}$ | $7.65^{+0.22}_{-0.24}$ | $0.1058^{+0.0008}_{-0.0008}$ | $0.040 \pm 0.003$ | $215^{+6}_{-5}$ | (3) |
| HAT-P-23 | $87.4^{+1.7}_{-2.0}$ | $4.26^{+0.13}_{-0.14}$ | $0.1113^{+0.0010}_{-0.0009}$ | $0.096 \pm 0.024$ | $121^{+11}_{-9}$ | (4) |
| WASP-10 | $88.26^{+0.26}_{-0.22}$ | $11.55^{+0.18}_{-0.17}$ | $0.1605^{+0.0011}_{-0.0012}$ | $0.0473 \pm 0.0032$ | $166^{+10}_{-9}$ | (5) |

(1) Consistent with Hartman et al. (2009) and Knutson et al. (2014).
(2) $e$ and $\omega$ set from Knutson et al. (2014) after their solution for other planets in the system.
(3) Solved using combined radial velocity data from Buchhave et al. (2010), Moutou et al. (2011) and Knutson et al. (2014).
(4) Solved using combined radial velocity data from Bakos et al. (2009) and Moutou et al. (2011).
(5) $e$ and $\omega$ set from Knutson et al. (2014) after their solution for a linear trend in the radial velocities.

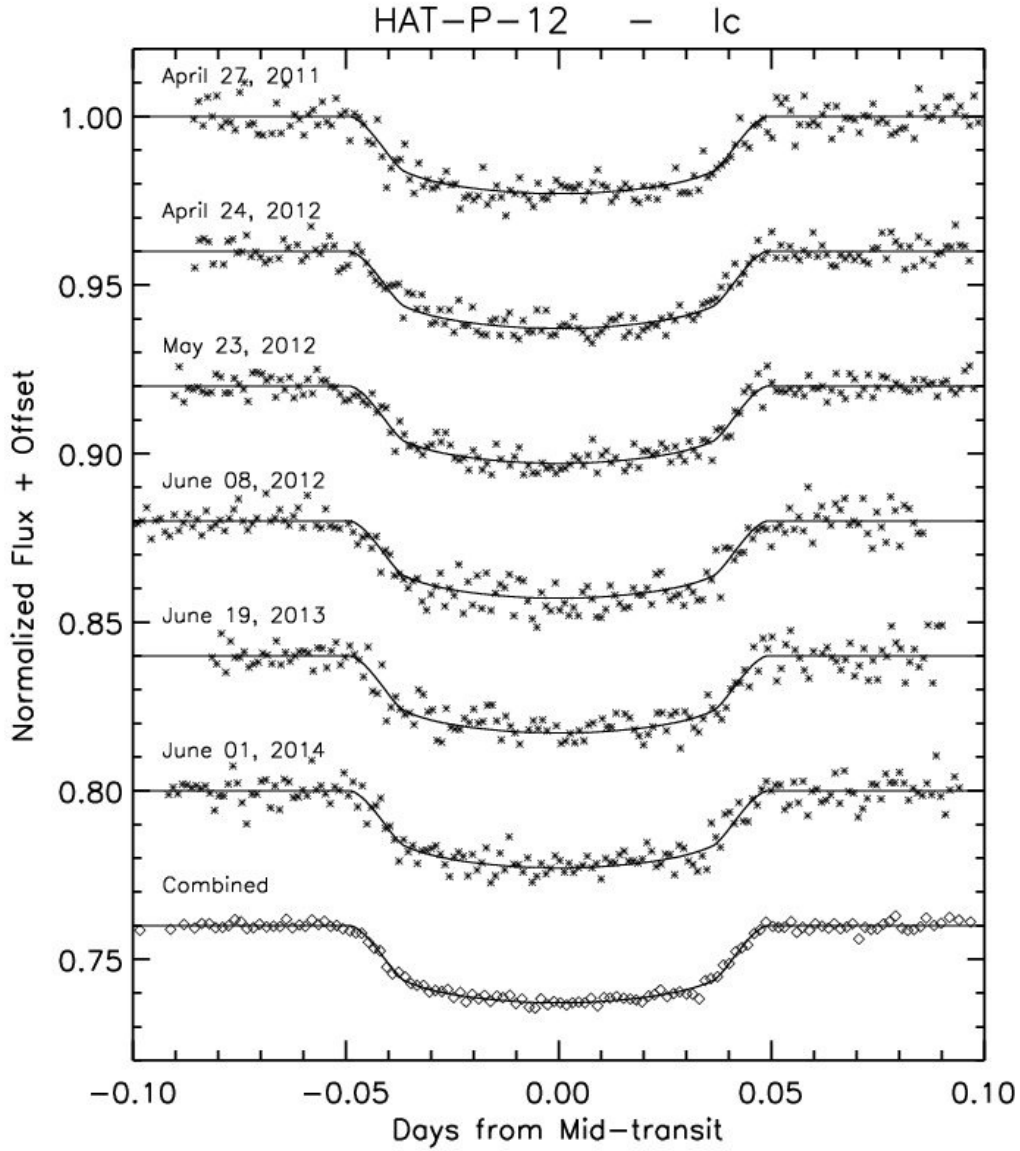

Figure 1a: Six transits of HAT-P-12b recorded at the Universidad de Monterrey Observatory through a Johnson-Cousins Ic filter are presented vertically offset for clarity. The resulting combined light curve is shown at the bottom. The individual photometric measurements are represented by asterisks, while the diamonds represent the bin averages in the combined light curve. The derived best-fit model is shown as a solid line for all cases. Average point-to-point variation values ranged between 0.0031 and 0.0045 for the individual light curves, while the value for the combined one was 0.0012.

Figure 1b: Residuals after a linear period fit to all available mid-transit times for HAT-P-12b. The diamonds represent the data available in the literature while the filled circles are the new transits presented here. The parameters used to derive this figure are P = 3.2130589 ± 0.0000003 days and $T_c$ = 2454419.19584 ± 0.00009 BJD_TDB.

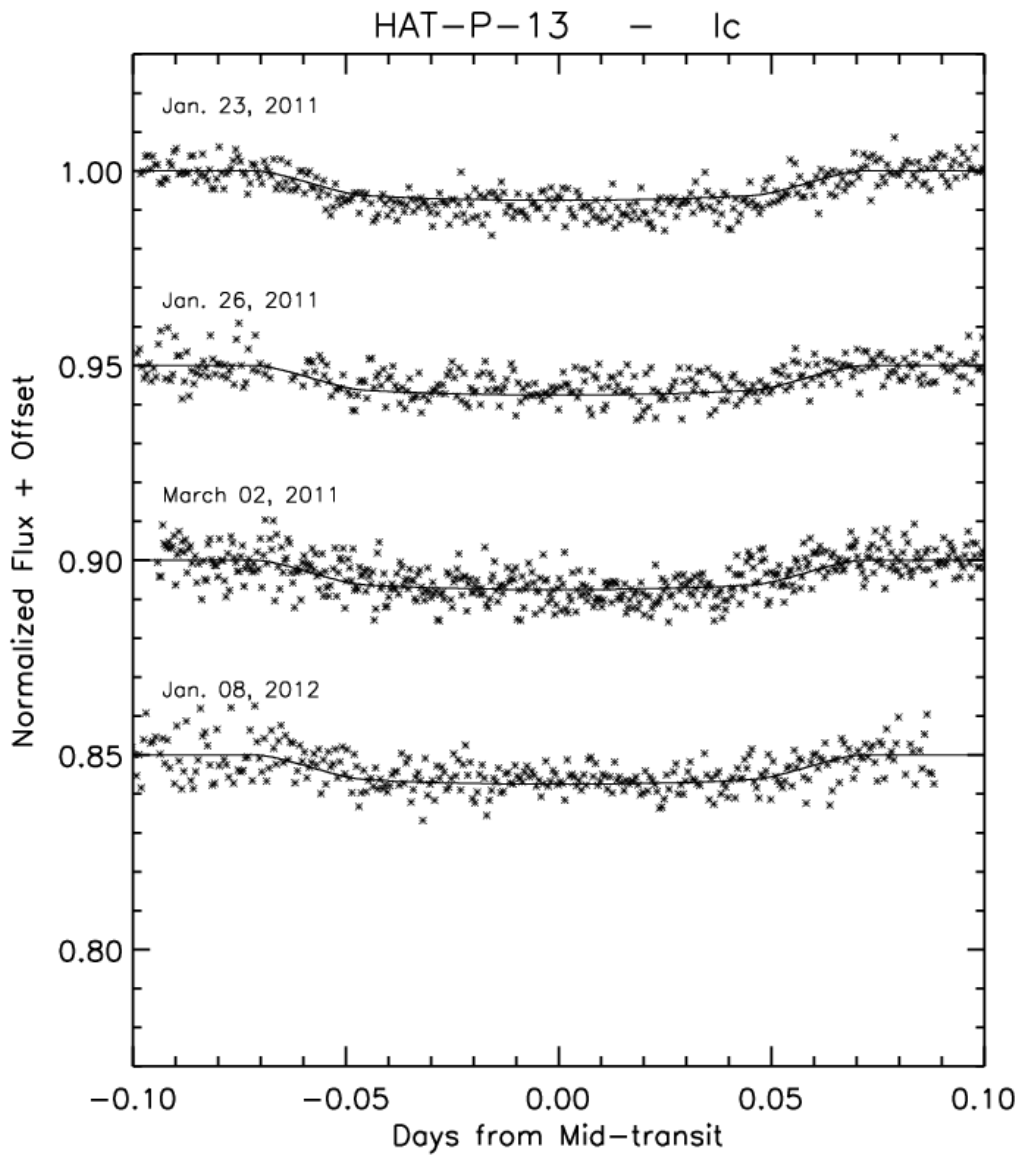

Figure 2a: Four transits of HAT-P-13b recorded at the Universidad de Monterrey Observatory through a Johnson-Cousins Ic filter are presented vertically offset for clarity. The asterisks are the individual photometric measurements. The best-fit model for the combined light curve is represented by the solid line. Average point-to-point variation values ranged between 0.0033 and 0.0043 for these particular individual light curves.

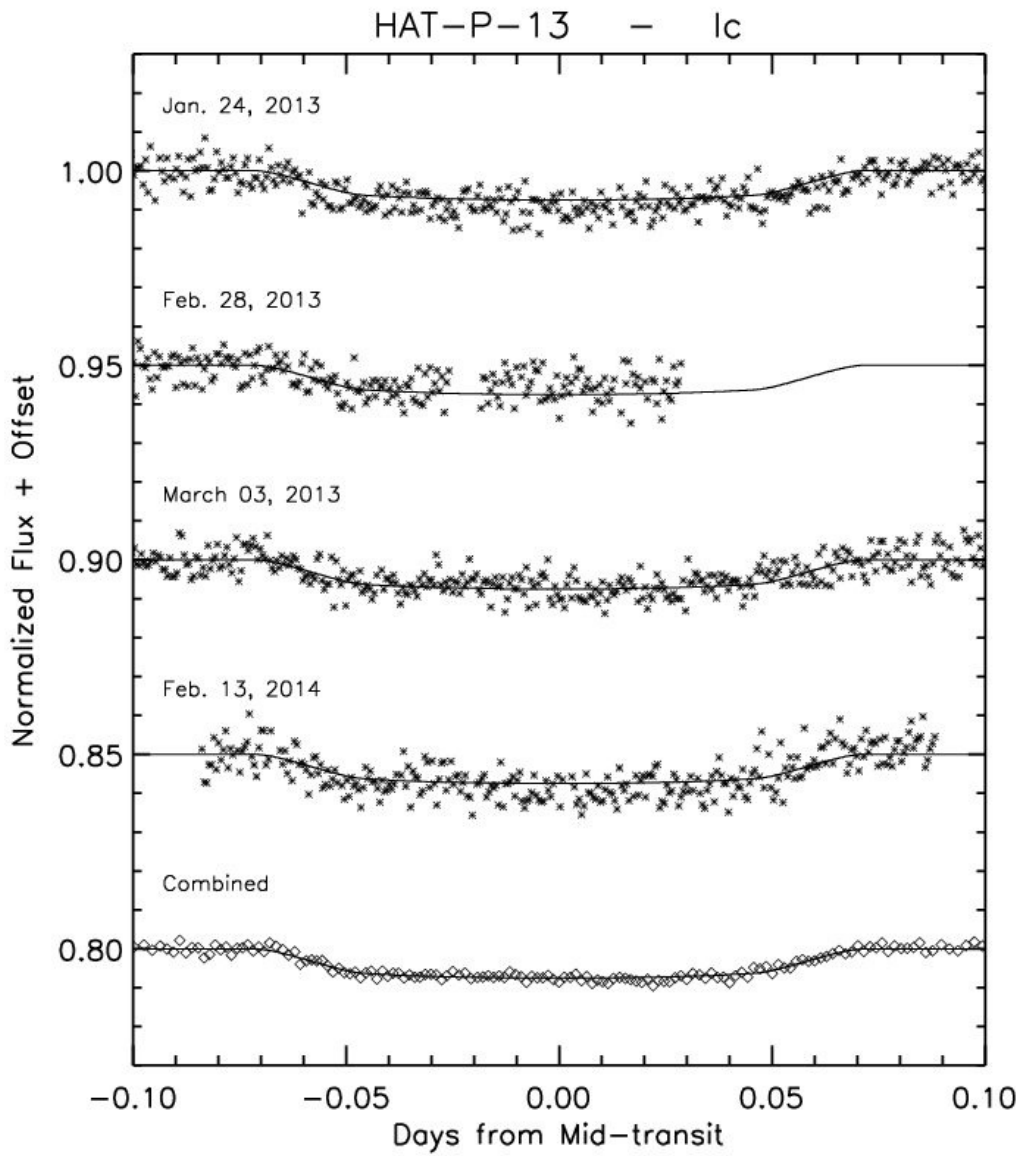

Figure 2b: Light curves of four more transits of HAT-P-13b registered at the Universidad de Monterrey Observatory. The 2013 February 28 transit was interrupted by clouds. The asterisks represent the individual measurements and the diamonds represent the bin averages after registering and combining the data for all eight transits of this system. The best-fit model to the combined light curve is represented by the solid line. Average point-to-point variation values ranged between 0.0032 and 0.0042 for these four individual light curves, while the value for the combined one was 0.0010.

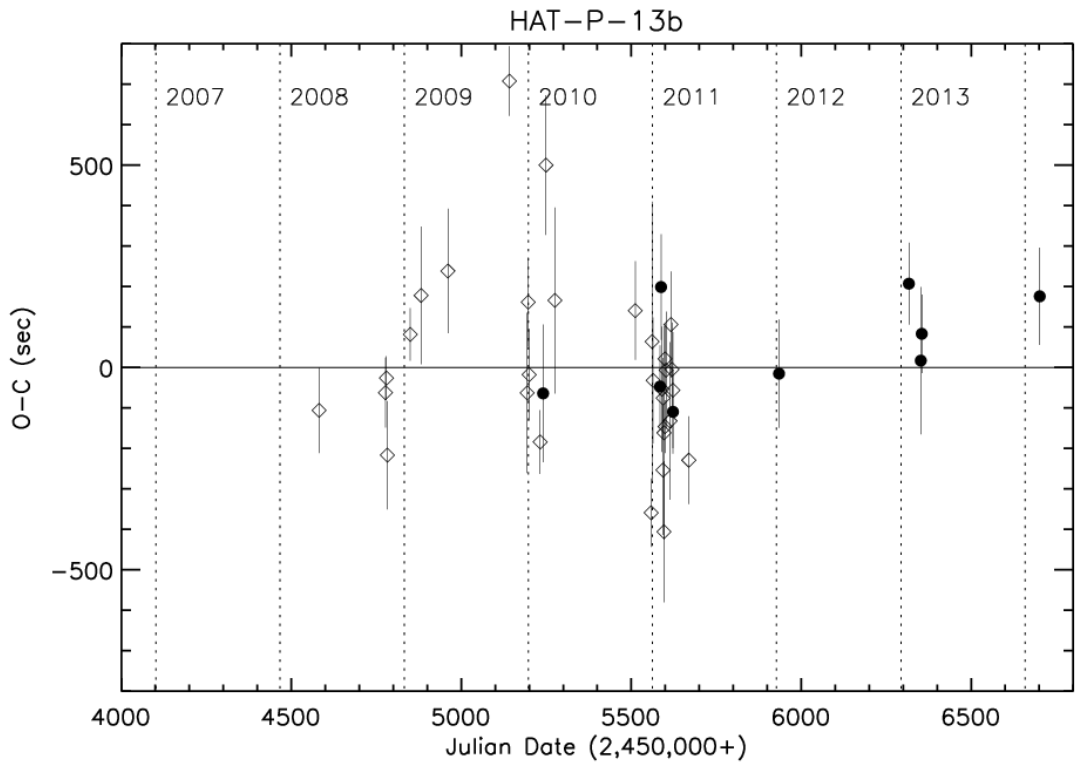

Figure 2c: Residuals after a linear period fit to the available mid-transit times for HAT-P-13b. The diamonds represent the data available in the literature (see the text) and the filled circles are the eight new transits presented in Figures 1a and 1b, plus an additional transit obtained on 2010 February 13 (see Table 1) through a Sloan z' filter that was not included in constructing the derived light curve. HAT-P-13 is known to contain at least another planet that interacts gravitationally with the transiting one, but its influence in the mid-transit times is too small to discern from the residuals because of the large uncertainties in the mid-transit times. The parameters used to derive this figure are P = 2.9162433 ± 0.0000012 days and $T_c$ = 2455176.53893 ± 0.00022 BJD_TDB.

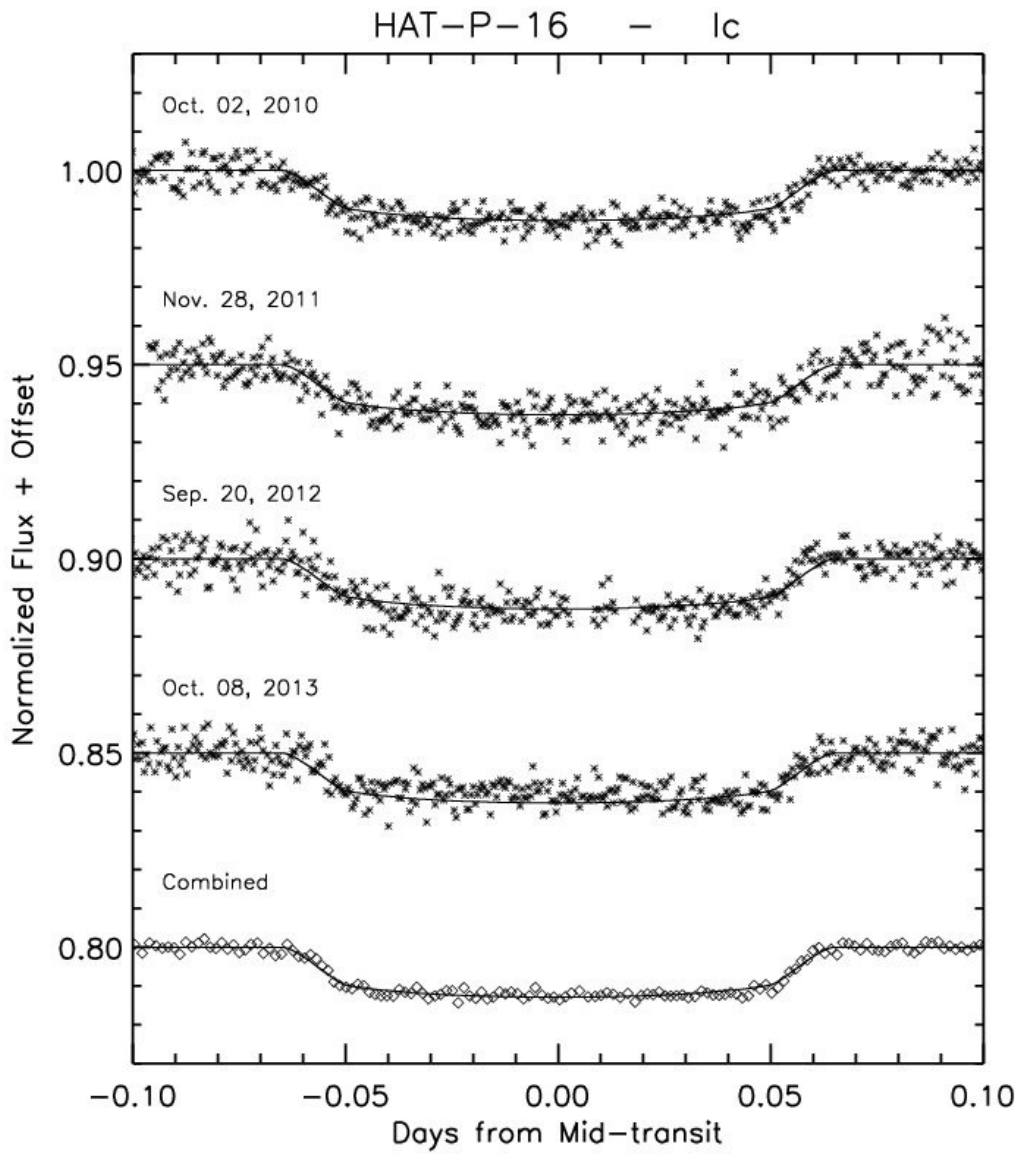

Figure 3a: Four transits of HAT-P-16b recorded at the Universidad de Monterrey Observatory through a Johnson-Cousins Ic filter are presented vertically offset for clarity. The resulting combined light curve is shown at the bottom. The individual photometric measurements are represented by asterisks, while the diamonds represent the bin averages in the combined light curve. The derived best-fit model is shown as a solid line for all cases. Average point-to-point variation values ranged between 0.0030 and 0.0042 for the individual light curves, while the value for the combined one was 0.0010.

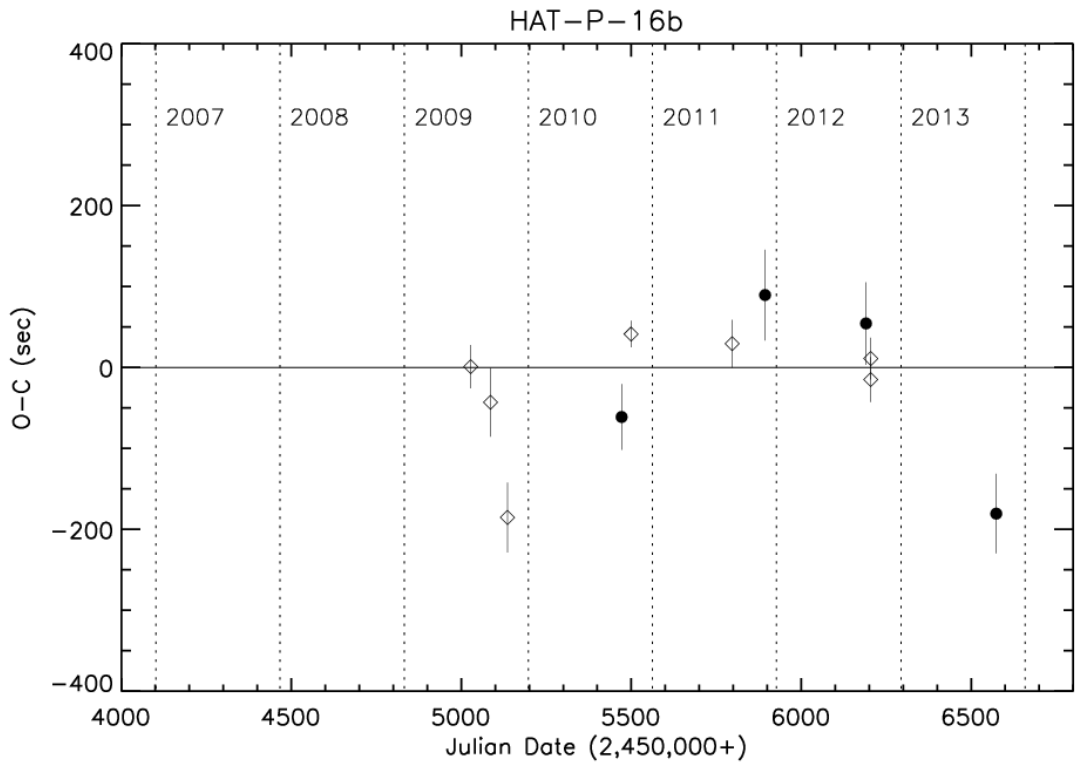

Figure 3b: Residuals after a linear fit to the period of HAT-P-16b. The solid circles are the mid-transit times presented in this paper, while the diamonds are the data from the literature. The parameters used to derive this figure are P = 2.7759704 ± 0.0000007 and $T_c$ = 5027.59292 ± 0.00019.

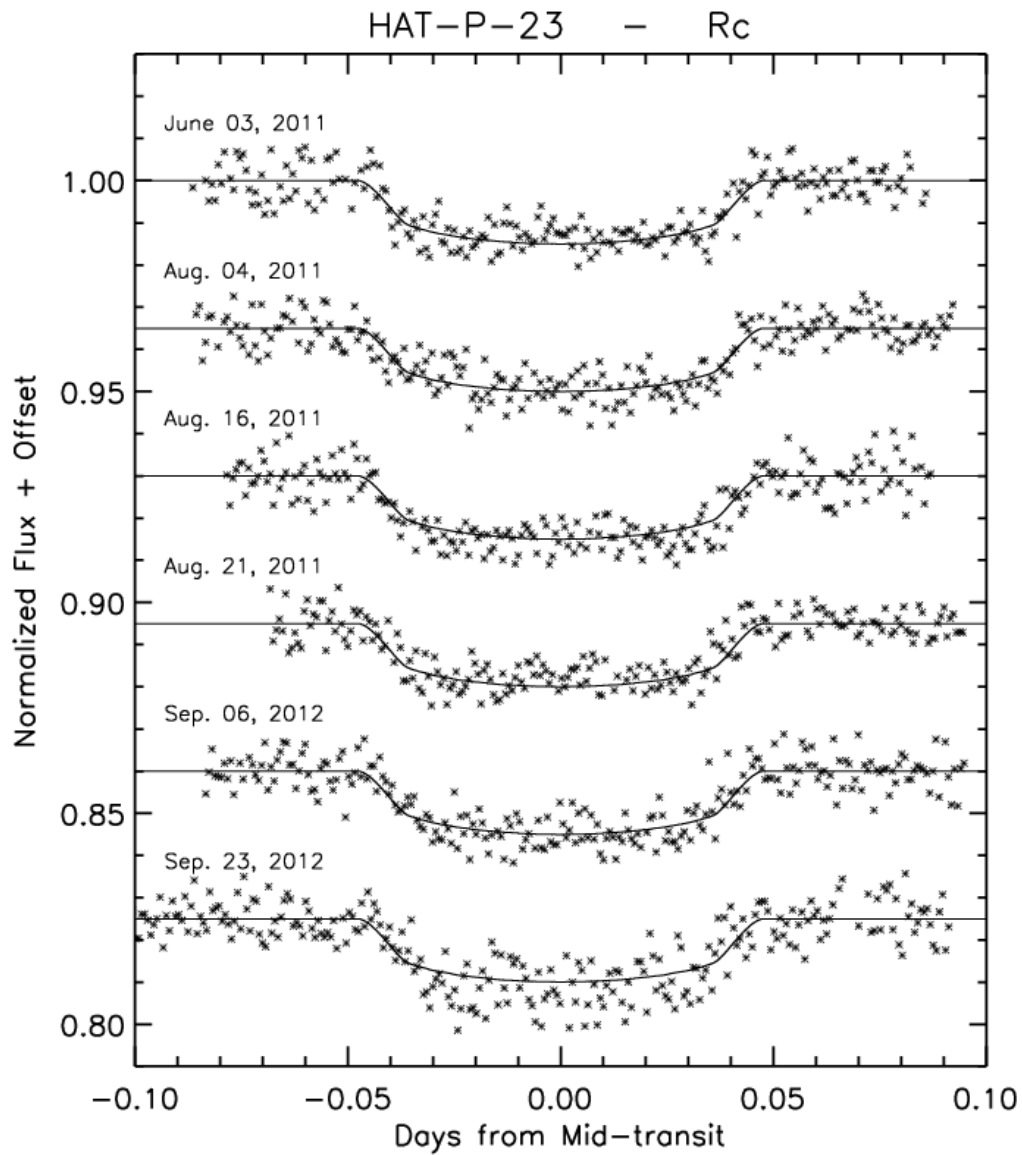

Figure 4a: Six transits of HAT-P-23b registered through a Johnson-Cousins Rc filter at the Universidad de Monterrey Observatory are presented vertically offset for clarity. The asterisks represent the individual photometric measurements and the solid line is the best-fit model to the combined light curve. Average point-to-point variation values ranged between 0.0038 and 0.0053 for these particular individual light curves.

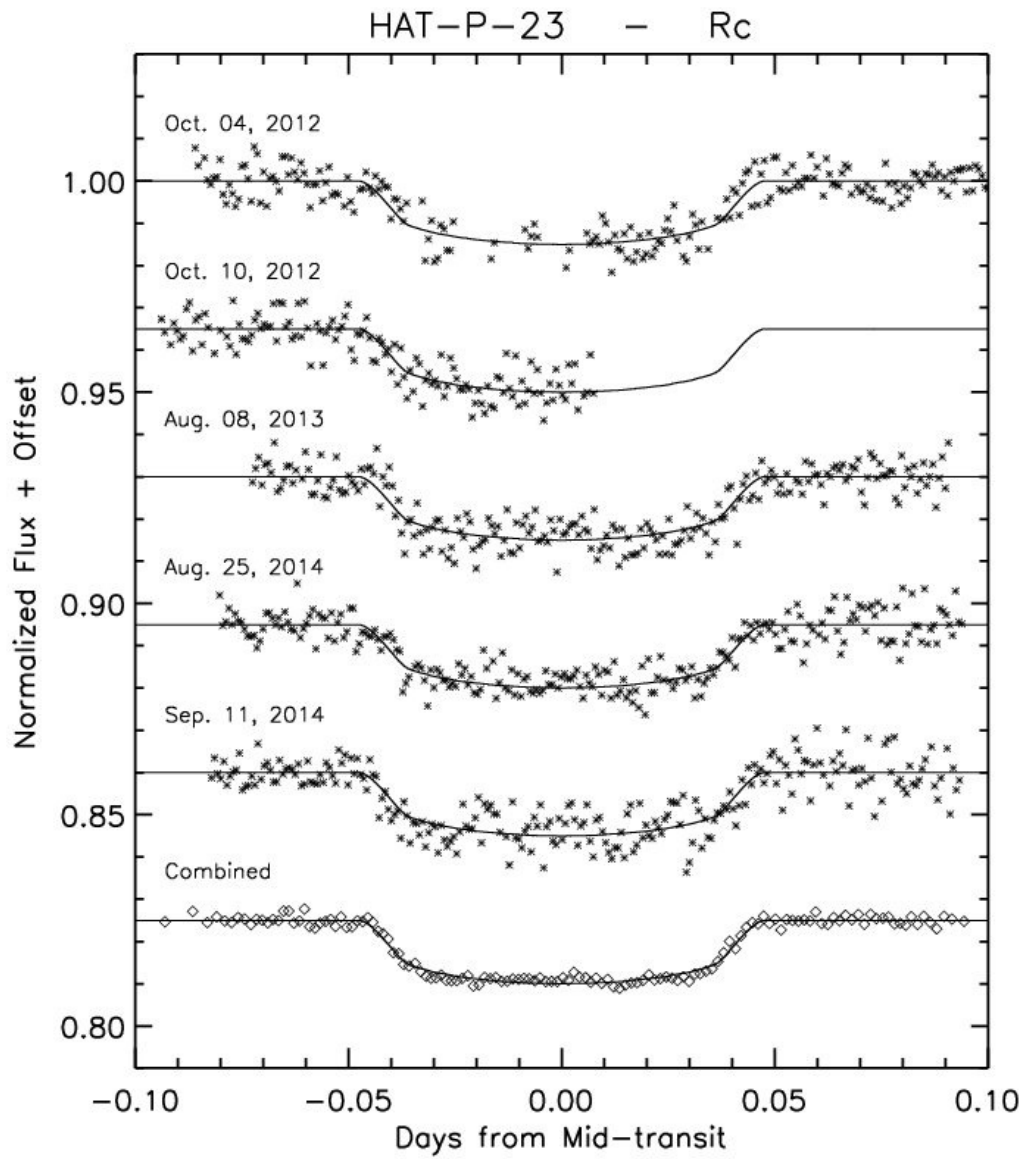

Figure 4b: Five more transits of HAT-P-23b recorded at the Universidad de Monterrey Observatory through a Johnson-Cousins Rc filter are presented vertically offset for clarity. The 2012 October 4 and 12 light curves were affected by clouds. The asterisks represent the individual photometric measurements. The bottom light curve, represented by diamonds, is the combination of all 11 individual light curves for this system. The solid line is the best model fit to the combined light curve. Average point-to-point variation values ranged between 0.0036 and 0.0046 for these five individual light curves, while the value for the combined one was 0.0010.

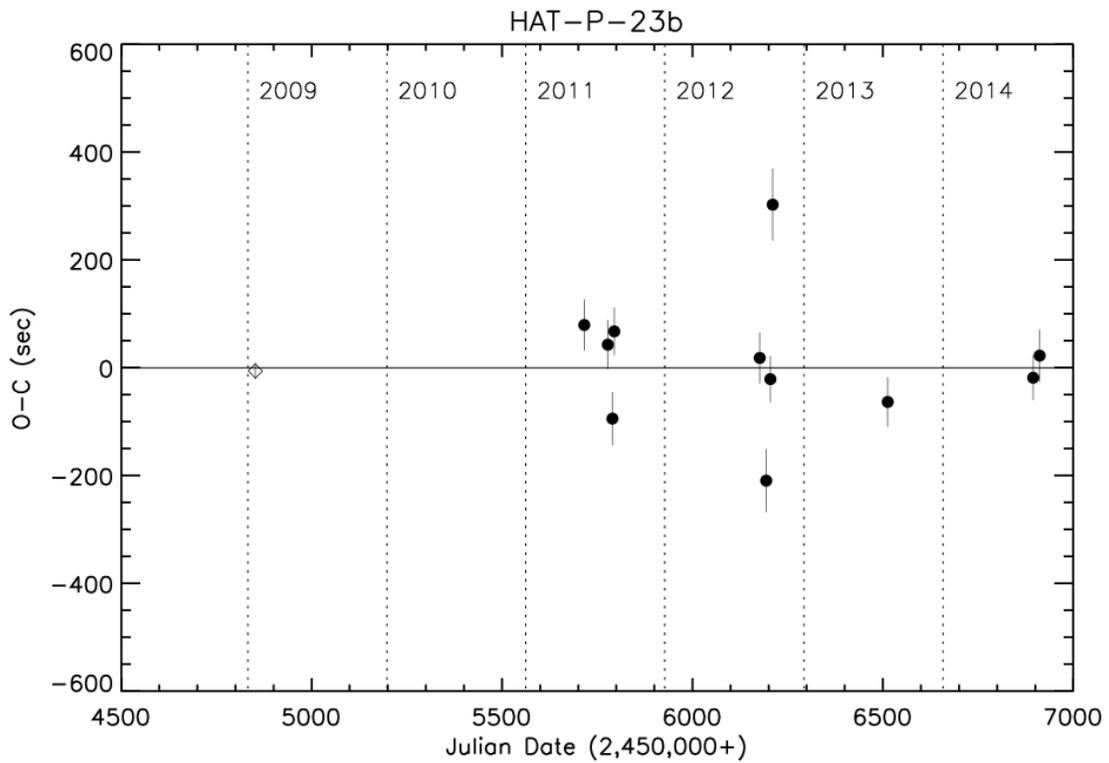

Figure 4c: Residuals after a linear period fit to the mid-transit times available for HAT-P-23b. The diamond represents the only ephemeris point from Bakos et al. (2011) available in the literature, and the filed circles are the mid-transit times of the data presented in Figs. 4a and 4b. The four transits observed in 2011 were those analyzed in Ramón-Fox & Sada (2013). The parameters used to derive this figure are P = 1.2128867 ± 0.0000002 days and $T_c$ = 4852.26548 ± 0.00017 BJD_TDB.

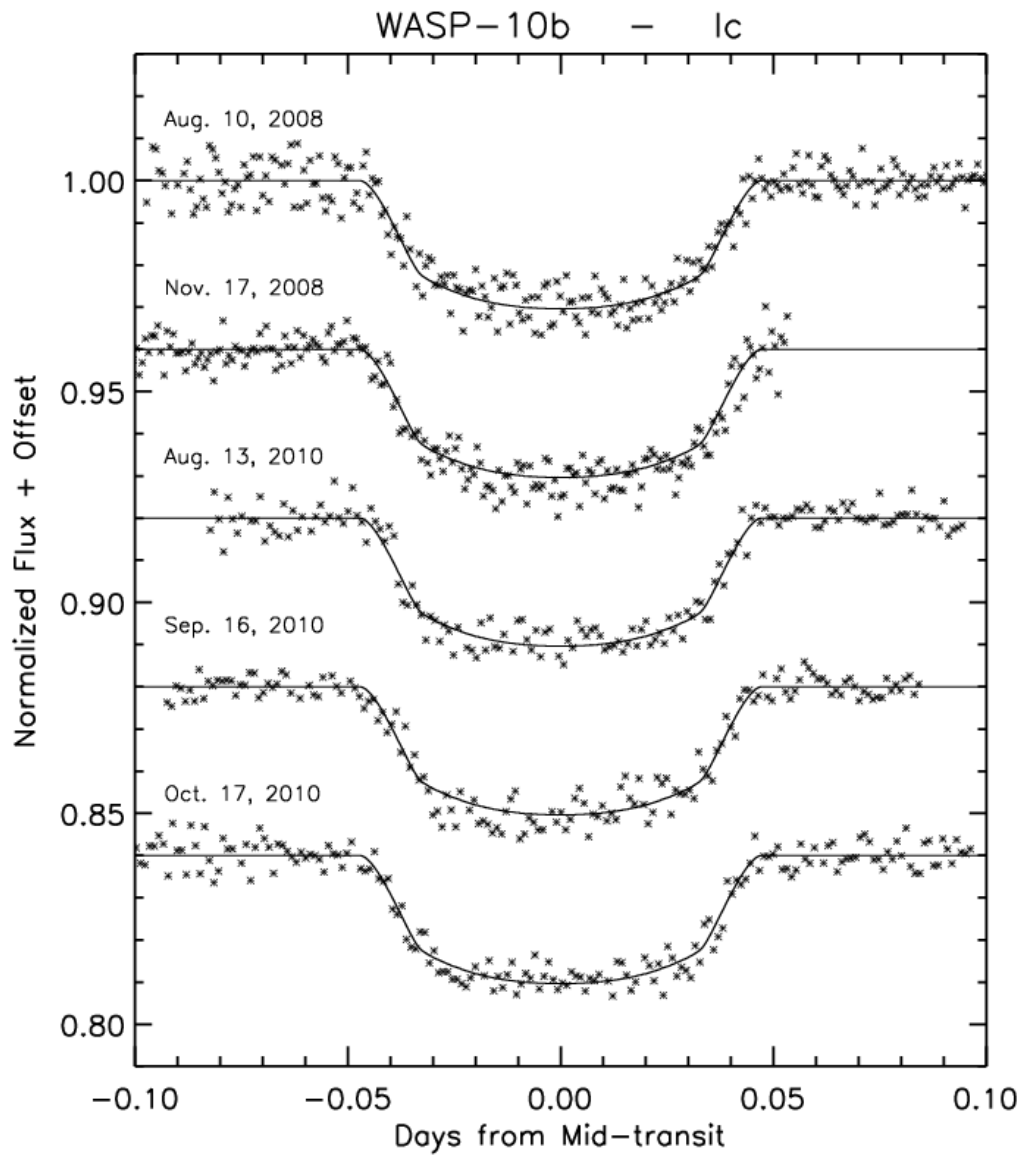

Figure 5a: Five transits of WASP-11b registered at the Universidad de Monterrey Observatory using a Johnson-Cousins Ic filter are presented vertically offset for clarity. The individual photometric measurements are represented by the asterisks, and the solid line is the best-fit model to the combined light curve shown in Fig 5b. Average point-to-point variation values ranged between 0.0031 and 0.0047 for these particular individual light curves.

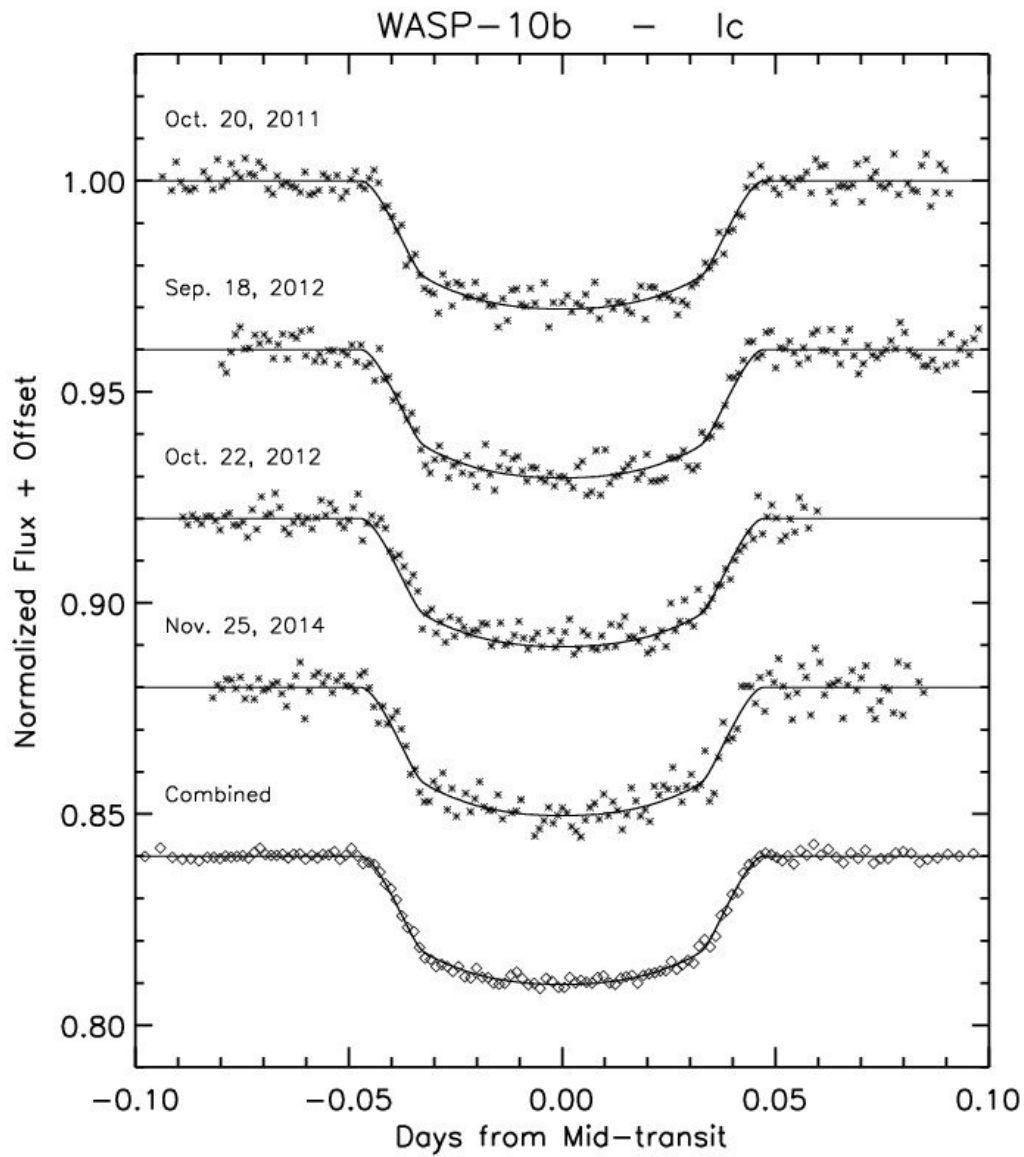

Figure 5b: Four more light curves of transits by WASP-10b recorded at the Universidad de Monterrey Observatory using a Johnson-Cousins Ic filter are presented vertically offset for clarity. The asterisks are the individual photometric measurements, and the diamonds on the bottom light curve represent the average of the bins after combining the data from all nine transits for this system. The solid line is the best-fit model to the combined transit light curve. Average point-to-point variation values ranged between 0.0031 and 0.0039 for these four individual light curves, while the value for the combined one was 0.0010.

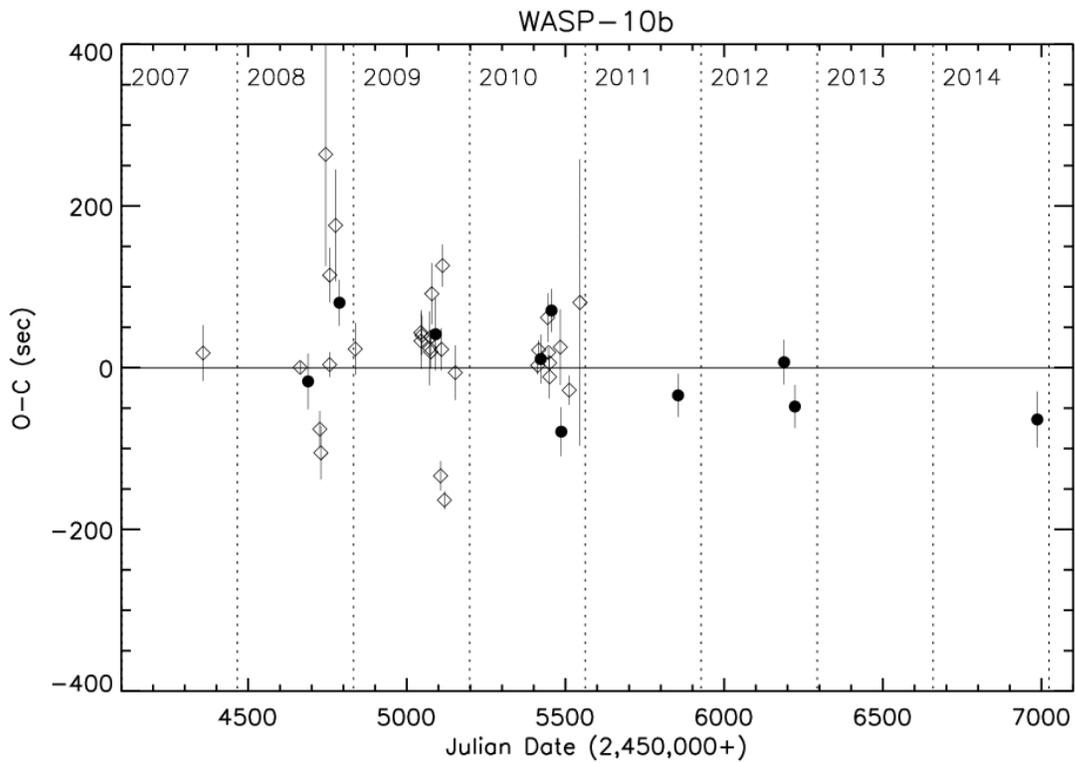

Figure 5c: Residuals after a linear period fit to all mid-transit times. The diamonds represent the data available in the literature, and the filled circles are the mid-transit times of the nine light curves presented in this work, plus another transit registered on 2009 September 16 through a different filter (see Table 1) but not included in the construction of the combined light curve. The parameters used to derive this figure are (P = 3.0927295 ± 0.0000003 days and $T_c$ = 4664.03804 ± 0.00006 BJD_TDB.

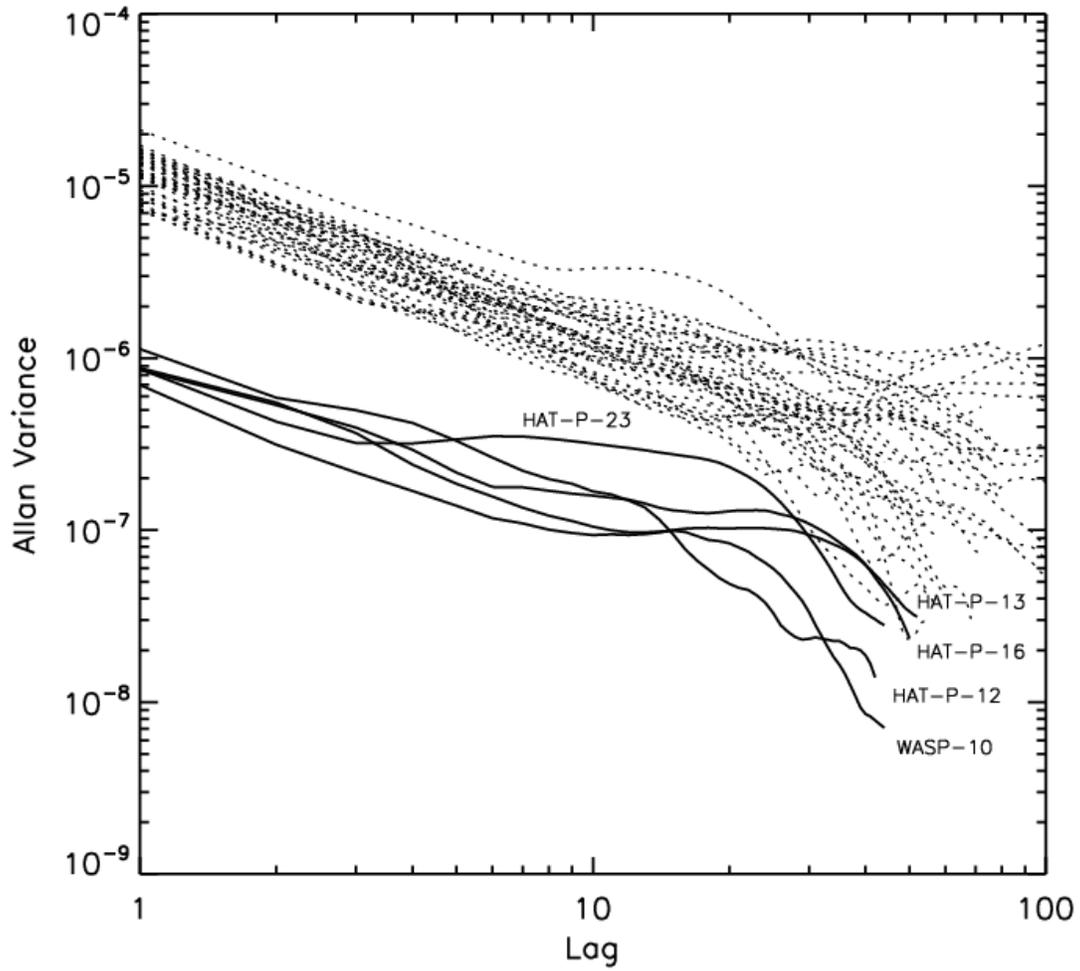

Figure 6: Allan variance of the residuals as a function of lag for the individual light curves (dotted lines) and combined light curves (labeled solid lines). The overall trend for both the individual and combined light curves corresponds to uncorrelated (white) noise. There is also some indication of flicker (pink) noise present for the combined light curves, in particular, for HAT-P-23. The Allan variance values for the combined light curves are about an order of magnitude smaller than those for the individual light curves.